\documentclass[twocolumn,preprintnumbers,superscriptaddress,prl]{revtex4}
\usepackage{graphicx}
\usepackage{times}
\usepackage{epsfig}
\usepackage{amsmath}
\usepackage{amsfonts}
\usepackage{amssymb}
\usepackage{url}
\usepackage{subfigure}
\newcommand{\be}{\begin{equation}}
\newcommand{\ee}{\end{equation}}
\newcommand{\bea}{\begin{eqnarray}}
\newcommand{\eea}{\end{eqnarray}}

\begin{document}
\pagestyle{plain}
\title{
Observable Gravity Waves From U(1)$_{B-L}$ Higgs 
and Coleman-Weinberg Inflation
}
\author{Nobuchika Okada}
\affiliation{
Department of Physics and Astronomy,
University of Alabama,
Tuscaloosa, AL 35487, USA
}
\author{Qaisar Shafi}
\affiliation{
Bartol Research Institute,
Department of Physics and Astronomy,
University of Delaware, Newark, DE 19716, USA
}


\begin{abstract}

We present a realistic non-supersymmetric inflation model 
 based on a gauged U(1)$_{B-L}$ symmetry and 
 a tree-level Higgs potential. 
The inflaton is identified with the scalar field 
 which spontaneously breaks U(1)$_{B-L}$, and 
 we include radiative corrections \`a la Coleman-Weinberg 
 in the inflaton potential. 
If the scalar spectral index $n_s$ lies close to 0.96, 
 as indicated by the recent Planck and WMAP 9-yr measurements, 
 the tensor-to-scalar ratio $r$, a canonical measure 
 for gravity waves, exceeds 0.01. 
Thus, according to this model, gravity waves should be found 
 in the near future. 
In this case, the quantity $|dn_s/d \ln k|$ 
 lies in the range $0.004-0.005$. 
Successful baryogenesis can be realized in this class of 
 models either via thermal or non-thermal leptogenesis.
 
\end{abstract}
\maketitle

The highly successful Standard Model (SM) of the strong, weak 
 and electromagnetic interactions possesses an accidental 
 global U(1)$_{B-L}$ symmetry at the renormalizable level. 
This symmetry can be upgraded to an anomaly free local 
 gauge symmetry by introducing three 
 SM singlet (right-handed) neutrinos.

Within the framework of supersymmetric hybrid 
 inflation~\cite{HI1, HI2}, the symmetry breaking scale 
 of U(1)$_{B-L}$ in a scenario with a renormalizable 
 superpotential and minimal Kahler potential is estimated 
 to be around $1-2 \times 10^{15}$ GeV~\cite{FHI1, FHI2}. 
This version of supersymmetric U(1)$_{B-L}$ inflation 
 also predicts that the tensor-to-scalar ratio $r$, 
 a canonical measure of primordial gravity waves, 
 lies many orders of magnitudes below the observable 
 capabilities ($r \gtrsim 0.02$) of Planck and 
 other contemporary measurements. 
On the other hand, within a more elaborate framework 
 with a non-minimal Kahler potential, the quantity $r$ 
 is found to lie in the observable range 
 with an appropriate choice of parameters~\cite{RSW}.

In this paper we implement inflation within the context 
 of the minimal $B-L$ extension of the SM, 
 which consists of the SM supplemented by three right-handed (RH) 
 neutrinos and a complex scalar carrying two units of $B-L$ charge 
 whose vacuum expectation value (VEV) spontaneously 
  breaks U(1)$_{B-L}$ to $Z_2$.
Associated with the $B-L$ gauge symmetry breaking, 
 the RH neutrinos acquire their Majorana mass 
 and the seesaw mechanism~\cite{seesaw} is automatically 
 implemented to yield the tiny neutrino mass.

The inflationary phase is linked to U(1)$_{B-L}$ breaking 
 and is driven by an appropriate 
 Higgs potential~\cite{HiggsPotentialInflation, HiggsPotentialInflation2}, 
 with radiative corrections arising from the inflaton couplings 
 to RH neutrinos, SM Higgs doublet and U(1)$_{B-L}$ vector 
 gauge boson also taken into account~\cite{SS}. 
By requiring that the scalar spectral index $n_s$ lies close 
 to 0.96, as determined by the recent Planck~\cite{Planck} and 
 WMAP 9-years~\cite{WMAP9} measurements, we are able to
 provide a lower bound $r \gtrsim 0.01$. 
Thus, according to this model of inflation, primordial 
 gravity waves should be found in the near future. 
Note that the U(1)$_{B-L}$ symmetry breaking scale 
 has transPlanckian value, but the cosmic strings 
 associated with the spontaneous breaking of U(1)$_{B-L}$ 
 are inflated away. 
For a somewhat different model of U(1)$_{B-L}$ inflation 
 involving non-minimal coupling to gravity, see Ref.~\cite{ORS-BL}.

The presence of RH neutrinos with direct couplings 
 to the inflaton is naturally compatible 
 with either thermal~\cite{thermalLG} or 
 non-thermal~\cite{non-thermalLG} leptogenesis.


We consider a minimal $B-L$ extension of the SM 
 based on the gauge group 
 SU(3)$_c \times$SU(2)$_L\times$U(1)$_Y\times$U(1)$_{B-L}$, 
 and the particle content are listed in Table~1.
Here, three generations of right-handed neutrinos ($\nu_R^i$)
 are introduced in order to make the model free 
 from all gauge and gravitational anomalies. 
The VEV of the SM singlet scalar ($\Phi$) breaks the U(1)$_{B-L}$ 
 gauge symmetry, and at the same time generates masses 
 for the right-handed neutrinos.

The Lagrangian relevant for the seesaw mechanism is given as 
\bea 
 {\cal L} \supset -Y_D^{ij} \overline{\nu_R^i} H^\dagger \ell_L^j  
- \frac{1}{2} Y_N^i \Phi \overline{\nu_R^{i c}} \nu_R^i 
+{\rm h.c.},  
\label{Yukawa}
\eea
 where the first term generates the Dirac neutrino mass term 
 after electroweak symmetry breaking, 
 while the right-handed neutrino Majorana mass term 
 is generated through the second term associated with 
 the $B-L$ gauge symmetry breaking. 
Without loss of generality, we work in a basis
 where the second term is diagonalized and 
 $Y_N^i$ are real and positive.

\begin{table}[t]
\begin{center}
\begin{tabular}{c|ccc|c}
            & SU(3)$_c$ & SU(2)$_L$ & U(1)$_Y$ & U(1)$_{B-L}$  \\
\hline
$ q_L^i $    & {\bf 3}   & {\bf 2}& $+1/6$ & $+1/3$  \\ 
$ u_R^i $    & {\bf 3} & {\bf 1}& $+2/3$ & $+1/3$  \\ 
$ d_R^i $    & {\bf 3} & {\bf 1}& $-1/3$ & $+1/3$  \\ 
\hline
$ \ell^i_L$    & {\bf 1} & {\bf 2}& $-1/2$ & $-1$  \\ 
$ \nu_R^i$   & {\bf 1} & {\bf 1}& $ 0$   & $-1$  \\ 
$ e_R^i  $   & {\bf 1} & {\bf 1}& $-1$   & $-1$  \\ 
\hline 
$ H$         & {\bf 1} & {\bf 2}& $-1/2$  &  $ 0$  \\ 
$ \Phi$      & {\bf 1} & {\bf 1}& $  0$  &  $+2$  \\ 
\end{tabular}
\end{center}
\caption{
Particle content. 
 In addition to the SM particle content, 
 there are three right-handed neutrinos $\nu_R^i$ 
 ($i=1,2,3$ denotes the generation index) 
 and a complex scalar $\Phi$. 
}
\end{table}


The tree level potential of the model is given by 
\bea 
 V_{\rm tree} = \lambda \left( 
  \Phi^\dagger \Phi - \frac{v_{\rm BL}^2}{2} \right)^2
  + \lambda_{\rm mix} (\Phi^\dagger \Phi) (H^\dagger H) 
+ V_H,   
\label{Vtree}
\eea
where $V_H$ is the SM Higgs potential. 
Here we assume that $\lambda_{\rm mix}$ is sufficiently small  
 (this is justified later), and so it can be ignored
 in our analysis for inflation.

In our numerical analysis, we employ the renormalization group 
 improved effective potential at the 1-loop level. 
We identify the inflaton with the real part of the $B-L$ Higgs field 
 ($\phi =\sqrt{2}\Re[\Phi]$), and parameterize 
 the effective potential in the leading-log approximation as 
\bea
V = \lambda
\left[ 
\frac{1}{4}\left( 
\phi^2 - v_{\rm BL}^2
\right)^2 + 
a \log \left[\frac{\phi}{v_{\rm BL}} \right] \phi^4
+V_0 \right], 
\label{Veff}
\eea
where $a= \frac{\beta_\lambda}{16 \pi^2 \lambda}$ with 
\bea 
\beta_\lambda &=& 
20 \lambda^2 + 2 \lambda_{\rm mix}^2+ 2 \lambda\left( 
 \sum_i (Y_N^i)^2-24 g_{BL}^2 \right) \nonumber \\
&+& 96 g_{BL}^4 - \sum_i (Y_N^i)^4. 
 \label{coeffC}
\eea
Here we have fixed the renormalization scale 
 to be $v_{\rm BL}$. 
In the presence of quantum corrections at the 1-loop level 
 ($a \neq 0$), the potential minimum is shifted 
 from its tree level location ($v_{\rm BL}$), 
 and a constant potential  energy density ($V_0$) is 
 suitably chosen so as to reproduce the observed (almost vanishing)  
 cosmological constant at the potential minimum. 
In our analysis, only three free parameters 
 $\{\lambda, v_{\rm BL}, a \}$ are involved.

The inflation takes place as the inflaton slowly rolls down 
 to the potential minimum from an initial VEV smaller than 
 the VEV at the potential minimum. 
The inflationary slow-roll parameters are given by
\bea
 \epsilon (\phi) &=& \frac{1}{2} m_P^2 
 \left( \frac{V^\prime}{V} \right)^2, 
\; \;  \eta (\phi) = m_P^2 \left( \frac{V''}{V} \right), \nonumber \\ 
\zeta^2 (\phi) &=& m_P^4 \left(
 \frac{V'  V'''}{V^2} \right), 
\eea
where $m_P = 2.4 \times 10^{18}$ GeV is the reduced Planck mass, 
 and a prime denotes a derivative with respect to $\phi$. 
The slow-roll approximation is valid as long as the conditions 
 $\epsilon \ll 1$, $|\eta| \ll 1$ and $\zeta^2 \ll 1$ hold. 
In this case, the scalar spectral index $n_{s}$, 
 the tensor-to-scalar ratio $r$, and the running of 
 the spectral index $\alpha \equiv \frac{d n_{s}}{d \ln k}$ 
 are given by
\bea
 n_s &\simeq& 1-6 \epsilon + 2 \eta, 
\; \; 
   r \simeq 16 \epsilon,  \nonumber \\ 
\alpha&\equiv & \frac{d n_{s}}{d \ln k} \simeq 16 \epsilon  \eta 
   - 24  \epsilon^2 - 2 \zeta^2. 
\label{reqn}
\eea
The number of e-folds after the comoving scale $k$ has 
 crossed the horizon is given by
\bea
N_k= \frac{1}{\sqrt{2} m_P} 
 \int_{\phi_{\rm e}}^{\phi_k}
 \frac{d \phi}{\sqrt{\epsilon(\phi)}}\hspace{0.3cm},
\label{Ne}
\eea
where $\phi_k$ is the field value at the comoving scale $k$, 
 and $\phi_e$ denotes the value of $\phi$ at
 the end of inflation, defined by 
 max$(\epsilon(\phi_e) , |\eta(\phi_e)|,\zeta^2(\phi_e)) = 1$.
The amplitude of the curvature perturbation 
 $\Delta_{\mathcal{R}}$ is given by 
\begin{equation} \label{Delta}
 \Delta_{\mathcal{R}}^2 = \left. \frac{V}{24\,\pi^2 \,
m_P^4\,\epsilon } 
 \right|_{k_0},
\end{equation}
which should satisfy 
 $\Delta_{\mathcal{R}}^2 = 2.215  \times 10^{-9}$ 
 from the Planck measurement~\cite{Planck} 
 with the pivot scale chosen at $k_0 = 0.05$ Mpc$^{-1}$.

\begin{figure}[ht]
\begin{center}
{\includegraphics[width=0.9\columnwidth]{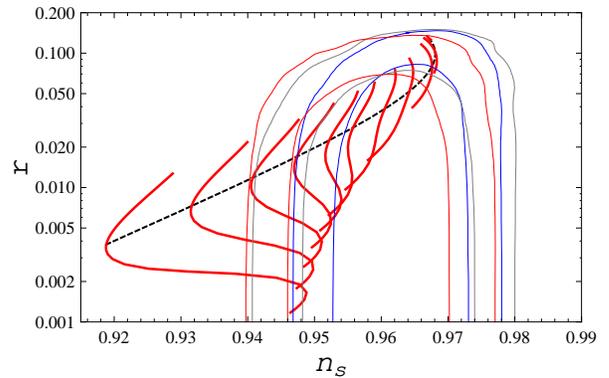}
\caption{
Predicted values of $n_s$ and $r$ 
 with various values of $v_{\rm BL}$ and $a$, 
 for $N_e = 60$, 
 along with the 68\% and 95\% CL contours 
 from the Planck measurement~\cite{Planck} 
 (Planck+WP: gray, Planck+WP+highL: red, Planck+WP+BAO: blue). 
Each thick red contour from left to right 
 corresponds to $v_{\rm BL}/m_P=10$, 
 $11$, $12$, $13$, $14$, $15$, $17$, $20$, $30$, $50$ and $500$, 
 respectively. 
On each contour for a fixed $v_{\rm BL}$, 
 the predicted values change from top to bottom 
 as $a$ varies in the range of $-0.2 \leq a \leq 1000$. 
The dashed line denotes the prediction with $a=0$ 
 for various $v_{\rm BL}$ values, from $v_{\rm BL}=10 \; m_P$
 to $500 \; m_P$. 
}
\label{ns-r}
}
\end{center}
\end{figure}

In our parameterization of the effective potential 
 of Eq.~(\ref{Veff}), the slow roll parameters 
 as well as the number of e-foldings only depend 
 on two parameters, $v_{\rm BL}$ and $a$. 
Thus, the predictions for $n_s$, $r$ and 
 $d n_{s}/d \ln k$ are given 
 for fixed values of $v_{\rm BL}$ and $a$, 
 while the quartic coupling constant $\lambda$ 
 is determined so as to satisfy 
 $\Delta_{\mathcal{R}}^2 = 2.215  \times 10^{-9}$. 
Fig.~\ref{ns-r} shows the predicted values of $n_s$ and $r$ 
 for various values of $v_{\rm BL}$ and $a$  
 with the number of e-foldings $N_e = 60$, 
 along with the 68\% and 95\% CL contours 
 from the Planck measurement~\cite{Planck}.  
Each thick red contour from left to right corresponds to 
 the results with $v_{\rm BL}/m_P=10$, 
 $11$, $12$, $13$, $14$, $15$, $17$, $20$, $30$, $50$ and $500$, 
 respectively. 
Along each contour for a fixed $v_{\rm BL}$, 
 the predicted values of $n_s$ and $r$ 
 change from top to bottom 
 with $a$ being varied from $a=-0.2$ to $a=1000$. 
The dashed line denotes the prediction with $a=0$ 
 for various $v_{\rm BL}$ values in the range of 
 $10 \; m_P \leq v_{\rm BL} \leq 500 \; m_P$. 
It is known~\cite{HiggsPotentialInflation2, LNW} 
 that in the limit $v_{\rm BL} \to \infty$, 
 the predictions coincide with those of the $m^2 \phi^2$ 
 chaotic inflation model, $(n_s, r) \simeq (0.967, 0.132)$.

\begin{figure}[ht]
\begin{center}
{\includegraphics[width=0.9\columnwidth]{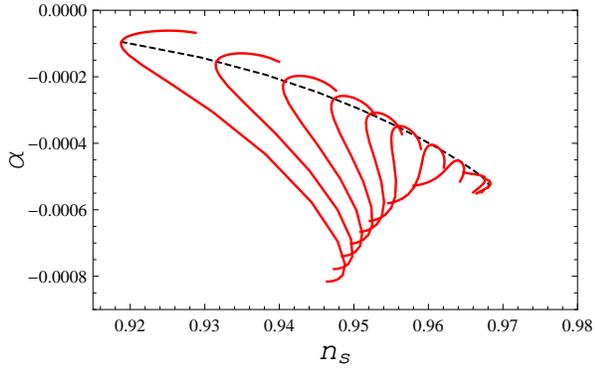}
\caption{
Predicted values of $\alpha \equiv d n_s/d \ln k$
 for various values of $v_{\rm BL}$ and $a$, 
 corresponding to Fig.~1.
}
\label{ns-alpha}
}
\end{center}
\end{figure}

The corresponding results for the predicted values of the running 
 of the spectral index $\alpha$ are depicted in Fig.~\ref{ns-alpha}. 
For the best fit value of $n_s \simeq 0.96$, 
 $r$ is predicted to exceed 0.01, so that 
 primordial gravity waves should be found in the near future. 
In this case, $-0.005 \lesssim \alpha \lesssim -0.004$, 
 which is consistent with the Planck measurement~\cite{Planck}, 
 $\alpha= -0.0134 \pm 0.018$.

For $a \gg 1$, the effective potential is dominated 
 by the 1-loop corrections, which corresponds to 
 the Coleman-Weinberg potential~\cite{CW, Shafi:2006cs} 
 and the B-L symmetry breaking occurs radiatively. 
In this case, we assume that the $B-L$ gauge coupling 
 ($g_{\rm BL}^4$ term) dominates in Eq.~(\ref{coeffC}). 
On the other hand, for $a < 0$, 
 we assume that the Yukawa coupling ($(Y_N^i)^4$ term) 
 dominates in Eq.~(\ref{coeffC}). 
These assumptions will be justified later. 
For $a <0$, the effective potential is unbounded from below 
 for $\phi \to \infty$, and thus we implicitly assume 
 that our universe sits at a local minimum. 
We find a lower bound $a \gtrsim -0.2$ 
 for the local minimum to exist.

After inflation is over, the inflaton decays to the SM particles 
 with subsequent thermalization of the universe. 
To discuss the post inflationary scenario, 
 we first calculate the mass spectrum of the model. 
In the Coleman-Weinberg limit ($ a \gg 1$), 
 we have found that for $v_{\rm BL} \gg m_P$, 
 the inflaton mass (calculated by the second derivative of 
 the effective potential at the minimum) 
 and the $B-L$ gauge boson mass are almost independent of $v_{\rm BL}$: 
\bea 
 && m_{\phi} \simeq 10^{13}\; {\rm GeV}. \nonumber \\  
 && m_{Z'} = 2 g_{\rm BL} \langle \phi \rangle \simeq 10^{17}\; {\rm GeV}, 
\eea
 where $\langle \phi \rangle \simeq v_{\rm BL}$ is 
 the inflaton VEV at the potential minimum. 
The mass spectrum is determined by the constraint 
 on $\Delta_R$, almost independently of $v_{\rm BL}$. 
Since $m_\phi \ll m_{Z'}$ and $\lambda \ll g_{\rm BL}^2$, 
 the assumption that the $B-L$ gauge coupling dominates 
 in Eq.~(\ref{coeffC}) is justified. 
As is well-known, this condition is necessary 
 for the Coleman-Weinberg mechanism for radiative symmetry breaking 
 to work~\cite{CW}. 

\begin{figure}[h,t]
\begin{center}
{\includegraphics[width=0.9\columnwidth]{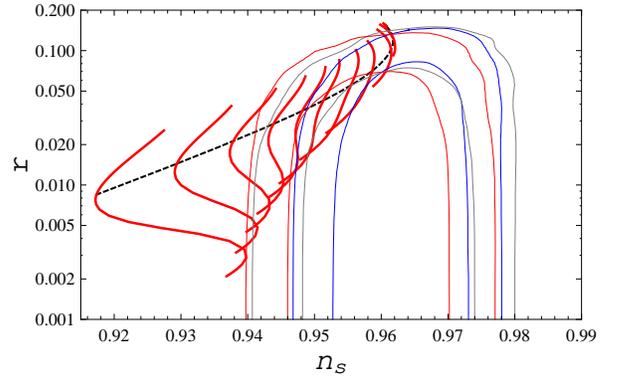}
\caption{
Same as Fig.~1 but for $N_e = 50$.
}
\label{ns-r-50}
}
\end{center}
\end{figure}

\begin{figure}[h,t]
\begin{center}
{\includegraphics[width=0.9\columnwidth]{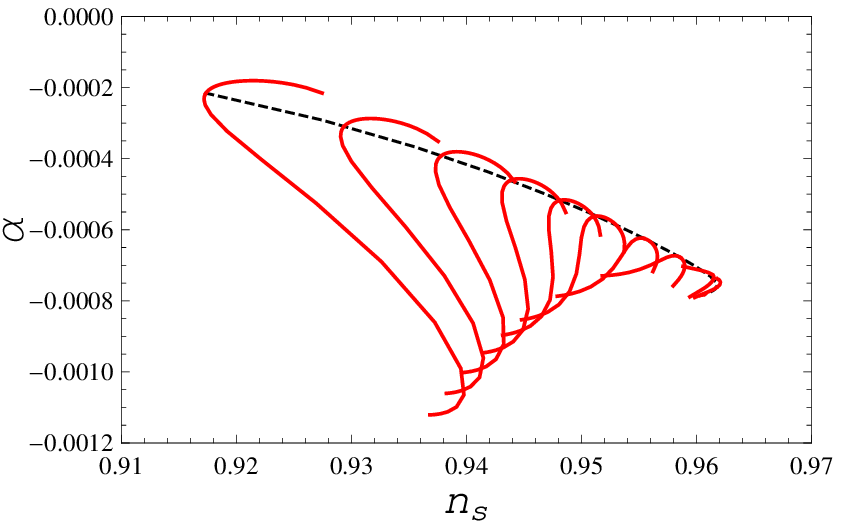}
\caption{
Same as Fig.~2 but for $N_e = 50$.
}
\label{ns-alpha-50}
}
\end{center}
\end{figure}

Next we estimate the reheat temperature 
 from the inflaton decay to a pair of SM Higgs bosons 
 through the coupling $\lambda_{\rm mix}$ in Eq.~(\ref{Vtree}). 
The decay width is given by 
\bea 
\Gamma (\phi \to hh) = 
 \frac{\lambda_{\rm mix}^2 \langle \phi \rangle^2}{32 \pi m_\phi}\hspace{0.3cm}.
\eea
In order for our calculation with the small decay width 
 approximation to be reliable, 
 we have a constraint on $\lambda_{\rm mix}$ 
 from the requirement that $\Gamma (\phi \to hh) \ll m_\phi$. 
The maximum value of $\lambda_{\rm mix}$ 
 is obtained via $\Gamma (\phi \to hh) = m_\phi$, 
 so that $\lambda_{\rm mix}^{\rm Max} 
 = \sqrt{32 \pi} m_\phi/\langle \phi \rangle $. 
A numerical analysis results in the relation 
 $ \lambda_{\rm mix}^{\rm Max} \sim g_{\rm BL}^2$
 for $10 \; m_P \leq v_{\rm BL} \leq 500 \; m_P$. 
Thus, $\lambda_{\rm mix} \ll \lambda_{\rm mix}^{\rm Max}$ 
 is negligible in Eq.~(\ref{coeffC}), 
 justifying the assumption.  
We estimate the reheat temperature by 
\bea 
\Gamma(\phi \to hh)= H= \sqrt{\frac{\pi^2}{90}g*}
 \frac{T_{\rm RH}^2}{m_P},     
\eea 
 and find $T_{\rm RH}^{\rm Max} \simeq 10^{15}$ GeV, 
 almost independently of $v_{\rm BL}$ 
 in the limit of $\lambda_{\rm mix}= \lambda_{\rm mix}^{\rm Max}$. 
This is the maximum value of the reheat temperature.

If the reheat temperature is sufficiently high
 for the right-handed Majorana neutrinos 
 to be in thermal equilibrium, 
 the thermal leptogenesis scenario~\cite{thermalLG} 
 works to reproduce the observed baryon asymmetry. 
We can choose appropriate values for $Y_N^i$ 
 to realize the condition for thermal leptogenesis to work~\cite{LGcond}, 
\bea  
 10^{10}\; {\rm GeV} \lesssim M_{N_i} < T_{\rm RH} \ll T_{\rm RH}^{\rm Max}, 
\eea
 where $M_{N_i}= Y_N^i \langle \phi \rangle/\sqrt{2}$
 is the Majorana mass of the RH neutrinos. 
Therefore, in our inflation scenario, 
 baryogenesis via (thermal) leptogenesis is successful.

When $\lambda_{\rm mix}$ is negligibly small,
 the inflaton dominantly decays to a pair of RH neutrinos 
 with a decay width 
\bea 
 \Gamma (\phi \to \nu_R^i \nu_R^i) 
 = \frac{(Y_N^i)^2}{64 \pi} m_\phi,   
\eea
 from which the reheat temperature is estimated as 
\bea 
 T_{\rm RH} \simeq 10^{14} \;{\rm GeV} \times Y_N^i.  
\eea
This reheat temperature is much smaller than 
 the RH neutrino mass $M_{N_i}=Y_N^i \langle \phi \rangle/\sqrt{2}$ 
 with a transPlanckian value of $\langle \phi \rangle$. 
In this case, we apply non-thermal leptogenesis~\cite{non-thermalLG} 
 to realize baryogenesis, where the RH neutrinos  produced 
 by the inflaton subsequently decay and create the observed baryon asymmetry 
 in the universe.

For the case with $a =-0.2$, we perform the same analysis 
 with the assumption that the Yukawa coupling ($(Y_N^i)^4$ term) 
 dominates in Eq.~(\ref{coeffC}). 
For simplicity, we assume $Y_N^1, Y_N^2  \ll Y_N^3$ 
 and then calculate the third generation RH neutrino
 mass as a function of $v_{\rm BL}$. 
We find the mass spectrum 
\bea
&& m_{\phi} \simeq 10^{13}\; {\rm GeV}, \nonumber \\  
&&  M_{N_3} = \frac{Y_N^3}{\sqrt{2}} \langle \phi \rangle 
 \simeq 10^{17}\; {\rm GeV}, 
\eea
 almost independently of $v_{\rm BL}$. 
This mass spectrum justifies the assumption 
 that the Yukawa coupling ($(Y_N^3)^4$ term) dominates 
 in Eq.~(\ref{coeffC}).

As in the Coleman-Weinberg limit, 
 we assume the inflaton dominantly decays to a pair 
 of the SM Higgs bosons and find 
 $\lambda_{\rm mix}^{\rm Max} \sim (Y_N^3)^2$, 
 so that $\lambda_{\rm mix} \ll \lambda_{\rm mix}^{\rm Max}$ 
 in Eq.~(\ref{coeffC}) is negligible, 
 justifying the assumption again. 
Since we have $T_{\rm RH}^{\rm Max} \simeq 10^{15}$ GeV, 
 the same as in the Coleman-Weinberg limit, 
 we can suitably arrange the value for $Y_N^i$ ($i=1,2$) 
 to realize the condition for thermal leptogenesis to work,
\bea  
 10^{10}\; {\rm GeV} \lesssim M_{N_1}, M_{N_2} < 
 T_{\rm RH} \ll T_{\rm RH}^{\rm Max} \ll M_{N_3}.
\eea
As previously discussed for the Coleman-Weinberg limit, 
 we may consider non-thermal leptogenesis 
 (for a successful baryogenesis) 
 if $\lambda_{\rm mix}$ is negligibly small 
 and the inflaton dominantly decays to the RH neutrinos  
 of the first and second generations.

For comparison, we also present the results 
 for the number of e-foldings $N_e=50$ 
 in Figs.~\ref{ns-r-50} and \ref{ns-alpha-50}. 
The predicted $r$ values are larger than those 
 for the $N_e=60$ case and a small portion of 
 the parameter space lies inside the 68\% CL contours 
 from the Planck measurement. 
We find that the resultant mass spectrum is quite similar 
 to the case for $N_e=60$ and the post inflationary 
 scenario is also essentially the same.

\begin{center}
{\bf Acknowledgments}
\end{center}
N.O. would like to thank the Particle Theory Group
 of the University of Delaware for hospitality during his visit.
This work is supported in part by the DOE Grants, 
 \# DE-FG02-10ER41714 (N.O.), and \# DE-FG02-12ER41808 (Q.S.).

%

\end{document}